# Superconductivity and metallic behavior in heavily doped bulk single crystal diamond and graphene/diamond heterostructure

Shisheng Lin,[1,2]* Xutao Yu,[1] Minhui Yang,[1] Huikai Zhong,[1] Jiarui Guo[1]

[1]*College of Information Science and Electronic Engineering, Zhejiang University, Hangzhou, 310027, P. R. China.*

[2]*State Key Laboratory of Modern Optical Instrumentation, Zhejiang University, Hangzhou, 310027, P. R. China.*

**Owing to extremely large band gap of 5.5 eV and high thermal conductivity, diamond is recognized as the most important semiconductor. The superconductivity of polycrystalline diamond has always been reported, but there are also many controversies over the existence of superconductivity in bulk single crystal diamond and it remains a question whether a metallic state exists for such a large band gap semiconductor. Herein, we realize a single crystal superconducting diamond with a Hall carrier concentration larger than $3\times10^{20}$ cm$^{-3}$ by co-doped of boron and nitrogen. Furthermore, we show that diamond can transform from superconducting to metallic state under similar carrier concentration with tuned carrier mobility degrading from 9.10 cm$^2$ V$^{-1}$ s$^{-1}$ or 5.30 cm$^2$ V$^{-1}$ s$^{-1}$ to 2.66 cm$^2$ V$^{-1}$ s$^{-1}$ or 1.34 cm$^2$ V$^{-1}$ s$^{-1}$. Through integrating graphene on a nitrogen and boron heavily co-doped diamond, the monolayer graphene can be superconducting through combining Andreev reflection and exciton mediated superconductivity, which may intrigue more interesting superconducting behavior of diamond heterostructure.**

Understanding the macroscopic quantum states of matter is critical for constructing novel electronic devices.[1-2] There is a rising interest in the superconductivity of carbon-based materials, as the twisted graphene provides a novel superconducting mechanism.[3-6] Similar with graphene, diamond is a carbon material existing in universe and under the earth, which can be synthesized under the environment of high temperature and pressure or by chemical vapor deposition (CVD) method enhanced by microwave plasma, which can dissociate $CH_4$ and $H_2$ into atomic carbon and hydrogen for assembling the single crystal diamond.[7] Diamond is also a very promising semiconductor which has a band gap larger than 5.5 eV with extremely hardness and thermal conductivity.[8] The study of the superconductivity of diamond originated from the polycrystalline superconducting sample of the heavily boron doped diamond by high temperature sintering, while the mechanism of superconductivity behavior in diamond is still unresolved and there are also some debates about the absence of superconductivity in bulk diamond single crystal.[8-10] The polycrystalline nature of diamond by high pressure method or CVD method brings many boundaries and even the second phase, which confuses the mechanism of superconductivity and limits the following improvements.[11-13] On the other hand, there are some efforts for understanding the metal-insulator transition of covalent diamond through the interaction between the valence band and impurity doping level and even superconducting transition temperature around 80 K is predicted.[14-17] However, the metallic behavior in bulk diamond under low temperature has rarely been explored although there are many reports of existence of metallic behavior in two dimensional (2D) materials.[18-22] Actually, there is also very few report of metallic behavior in bulk materials, where the metallic behavior may be highly correlated with the superconducting mechanism.[22-25] On the other hand, although an anomalous metallic state was observed during the superconductor-to-insulator transition in 2D electronic systems, but its properties are still unclear and poses great challenges to the traditional theories which expect no intermediate metallic phase.[26-28]

Superconductivity in 2D materials, such as the high-profile twisted bilayer graphene, has revealed many intriguing physical properties.[29,30] The limitations of phonons as mediators of electron pairing for realizing high critical temperature ($T_C$) superconductors motivate the search for novel superconducting mechanisms. However, the investigation on the superconducting behavior of 2D materials/superconductor heterostructure is quite limited, although where the Bose-Fermi hybrid system may permit the pairing of electron through exciton or virtual exciton.[31] Herein, by co-doped boron with nitrogen, single crystal bulk diamond with ultrahigh carrier concentration exceeding $3×10^{20}$ cm$^{-3}$, which equals to $3×10^{14}$ cm$^{-2}$ in 2D carrier concentration, has been realized. We show the tuned mobility of boron and nitrogen co-doped diamond permits the existence of metallic diamond as the mobility decreases from a superconducting state. By forming graphene/diamond heterostructure with the ultra-heavily co-doped diamond, we realized the superconducting graphene. Interestingly, the resistivity of graphene is not always zero while it crosses the zero line, which should be correlated with the Andreev reflection and exciton mediated superconductivity at the graphene/diamond interface.

**Results and Discussions**

Fig. 1a and Fig. 1b show the superconductivity behavior of diamond as revealed by resistivity and diamagnetism measurements. Fig. 1a shows the dependence of resistivity of diamond on temperature, which drops down to zero around 3 K, indicating that a supercurrent can persist without attenuation in the superconducting state, which was further demonstrated by the resistivity-temperature test on another diamond sample (Supplementary Fig. S1). The magnetization properties were measured by a superconducting quantum interference device down to 2 K. As shown in Fig. 1b, the splitting of the zero-field cooling (ZFC) and field-cooling (FC) curves at the $T_C$ indicates the superconducting property of the diamond, as the large curvature difference of the two curves below the $T_C$ originates from the large flux pinning force in the FC condition. The inset of Fig. 1b shows the magnetization versus magnetic field (M-H) curve measured at 2 K, where the large symmetric hysteresis curvature associates with the mixed state in type-II superconductors (Supplementary Fig. S2). According to the hysteresis phenomenon from the magnetization curve, the critical current density $J_C$ can be estimated by the assumption of a critical-state model with a formula $J_C = 30×\Delta M/d$, where $d$ is the size of the diamond sample ( 4.5 mm×4 mm× 37 μm), thus the $J_C$ at 0 T was estimated to be 226 A/cm$^2$. Fig. 1c and Fig. 1d show the X-ray diffraction (XRD) and Raman spectrum of boron and nitrogen co-doped diamond, respectively, which demonstrate the single crystal nature of those heavily doped diamond. In Fig. 1c, the XRD peak around 2θ =120° can be split into two peaks of 119.43° and 119.92°, due to the excitation of Cu K$_α$ at 0.154 nm and Cu K$_β$ line at 0.139 nm, respectively. The standard distance between (400) planes is 0.898 nm with XRD peak at 118.1° under the excitation of Cu K$_α$ at 0.154 nm, which is smaller than 119.43°, meaning that the lattice is shrank due to very heavy doping level of boron. Similar results are also shown in Supplementary Fig. S3, where XRD testing shows good homogeneity of the samples. In Fig. 1d and Supplementary Fig. S4, the Raman peaks at 1327.7 cm$^{−1}$ and 1326.6 cm$^{−1}$ are attributed to the in-plane vibration of sp$^3$ hybridized carbon atoms and internal stress during the growth process, resulting in a red shift from the intrinsic peak around 1332 cm$^{-1}$. Through Hall-effect measurements by constructing ohmic electrode at the four corners of diamond layer, the carrier concentration is normally larger than $2×10^{20}$ cm$^{-3}$.

Phase transition can change the properties of a material without the introduction of any additional atoms, so quantum phase transition has been a hot topic in condensed matter physics and materials in this century, herein the tailoring and understanding of the phase transitions in diamond is of major interest for both fundamental physics and applications. Since the discovery of diamond superconductivity, the existence and formation mechanism of quantum metallic phase in diamond have been an unresolved physical problem. We have tuned the carrier concentration and mobility through finely tuning the growth condition, we have found a finite resistivity remains while the phase transition happens at low temperature similar with the otherwise superconducting transition point. The clearly difference from the superconducting phase as shown in Fig. 1e and Fig. 1f suggests a significantly different quantum interference exists in this heavy doped diamond. As we can see from Fig. 1f, the resistivity of diamond no longer drops down to zero around the transition temperature 3 K, remains a value of 0.004 Ohms. The resistivity of diamond undergoes a significant change in magnitude, resulting in a transformation of its material properties, which indicates the occurrence of a transition to a metallic state (more experimental data can be seen in Supplementary Fig. S5).

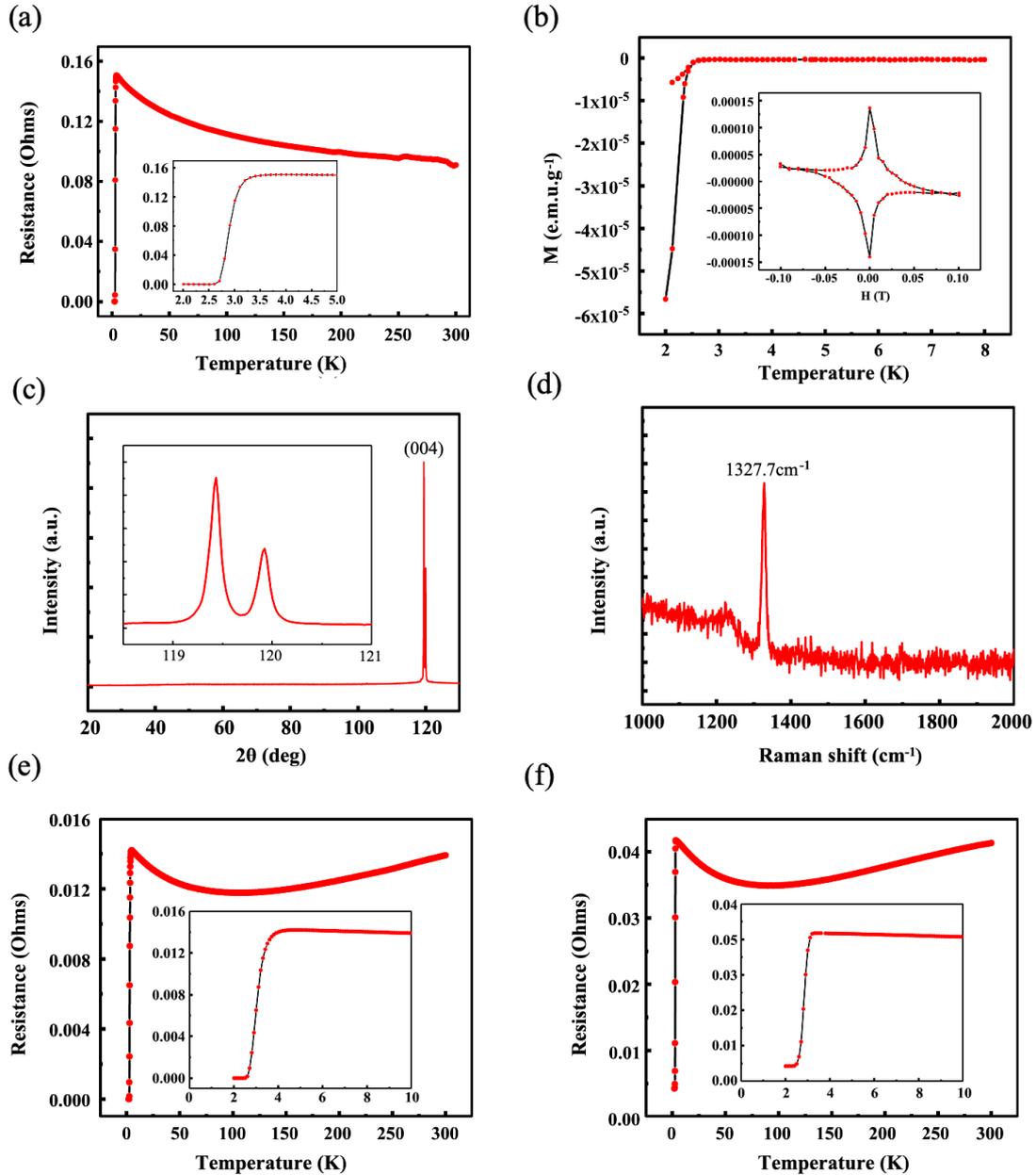

**Fig. 1 | Physical properties of superconducting heavily doped diamond. a,** typical resistivity as a function of temperature in superconductive diamond bulk layer. **b,** diamagnetism test of diamond bulk of size 4.5 mm×4 mm×37 μm. **c** and **d,** XRD spectrum and typical Raman spectrum of typical heavily doped single crystal diamond of size 7 mm×7 mm×0.1 mm, respectively. **e** and **f,** dependence of resistivity on the temperature for superconductive and metallic bulk single crystal diamond, respectively.

In order to gain a deeper understanding of the underlying physics of this metallic phase, we conduct a series of Hall-effect measurements. As shown in Table 1, the carrier concentration of the diamond samples is normally larger than 2×10$^{20}$ cm$^{-3}$. Compared with the superconducting samples D121313 and D010708, the metallic samples D120709 and D020405 show higher resistivity and hole concentration, however, much lower carrier mobility. Two mechanisms should be correlated with this phenomenon, one is the loss of superconducting pairing amplitude under a fermionic description and the second is the boson localization, where the boson may be caused by copper pair or exciton suggested by recent theories.[18,32,33] For all of the samples listed in Table 1, there are many nitrogen and boron atoms incorporated into the diamond lattice, which should induce the localization of electron and hole and subsequently form the excitons with residue holes as the boron doping concentration is times of that of nitrogen. A physical picture can be drawn that high mobility holes can be thoroughly paired through phonons or excitons, then superconductivity can be realized. If the mobility is not high enough, the holes is not paired thoroughly, leading to a metallic state. The superconducting state is different from the metallic state as the hole mobility is high enough to connect all those excitons whereas the metallic state has a degraded hole mobility. The physical image can be well fitted with the 2D superconductor in the presence of disorder. In our samples, the disorder can be finely described by Hall-effect measurements where the mobility is highly correlated with the disorder benefiting from the fact that diamond is a semiconductor.

**Table 1 | Different diamond samples and their parameters.**

| Sample | Thickness (μm) | Resistivity (Ω cm) | Carrier mobility (cm² V$^{-1}$s$^{-1}$) | Hole concentration (cm$^{-3}$) | State |
|---|---|---|---|---|---|
| D121313 | 37 | 2.07×10$^{-3}$ | 9.10 | 3.32×10$^{20}$ | Superconducting |
| D010708 | 285 | 5.80×10$^{-3}$ | 5.30 | 2.03×10$^{20}$ | Superconducting |
| D120709 | 289 | 6.12×10$^{-3}$ | 2.66 | 3.83×10$^{20}$ | Metallic |
| D020405 | 394 | 6.83×10$^{-3}$ | 1.34 | 6.83×10$^{20}$ | Metallic |

Fig. 2a shows the typical low-magnification transmission electron microscope (TEM) image of diamond, which demonstrates the uniform morphology and contrast. Fig. 2b shows the high-resolution TEM image of diamond and the inset shows the lattice fringe of 0.2037 nm, consistent with the distance of (110) plane of diamond. Fig. 2c shows the fast Fourier transform (FFT) of the Fig. 2b, where the electron incident direction is confirmed to be <$\bar{1}$01>. Fig. 2d, Fig. 2e and Fig. 2f show the energy dispersive spectrometer (EDS) mapping of boron, carbon and nitrogen elements in the diamond film, respectively.[15,16] The uniformly distribution of boron, carbon and nitrogen in diamond demonstrates there is no other phases rather than diamond in the bulk crystal. Fig. 2g, 2h, and 2i sequentially show the diffraction patterns of the three regions labeled from left to right in Fig. 2a, where the diffraction patterns are the same along the crystal, further demonstrating its nature of single crystal even under such a heavy boron doping level.

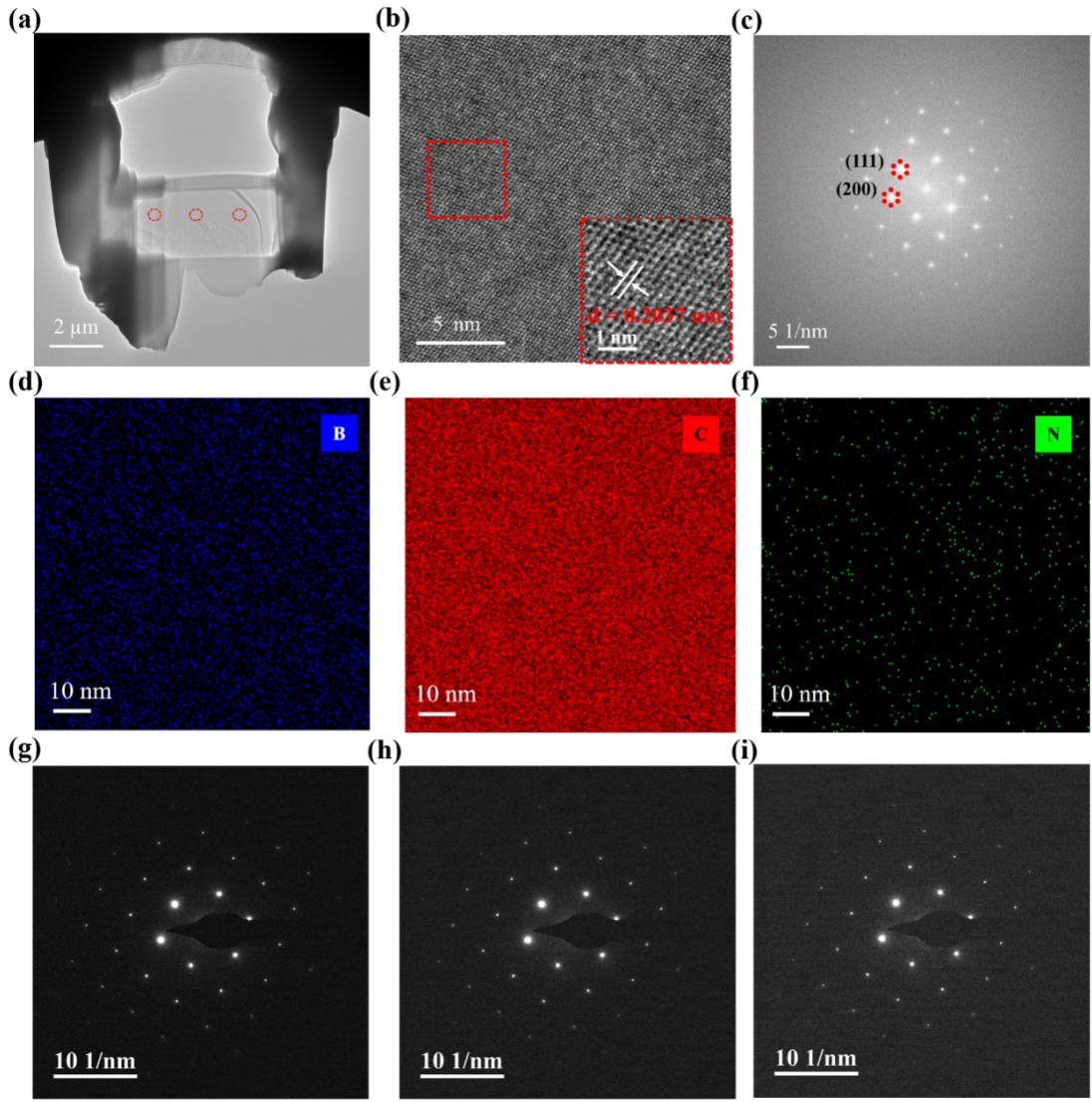

**Fig. 2 | The TEM characterization of diamond. a,** TEM image of heavily co-doped diamond. **b,** High-resolution TEM image shows the good crystal quality with a lattice fringe of 0.2037 nm. **c,** the FFT result of **b**. **d, e** and **f,** the distribution of boron, carbon and nitrogen in the diamond film. **g, h** and **i,** diffraction pattern of lattice structure along the <$\bar{1}01$> direction in different area to show the consistency of crystal structure.

The superconducting transition temperature changes as the applied magnetic field increases. Fig. 3a shows the temperature dependence of resistivity of the film under several values of magnetic fields up to 4 T for superconducting sample. The field dependence of the onsets of $T_C$ is plotted in Fig. 3b, the red dashed line is a linear fit to the onset values of $T_C$ over temperature with an equation can be expressed as $H =$

-1.3×$K$ + 5.7, where the $K$ stands for temperature. As we can see from Fig. 3b, the $T_C$ onset can approach the value of 5.7 T at absolutely zero kelvin temperature. The superconductivity behavior of the heavily boron and nitrogen co-doped diamond has been repeatably produced. It is argued that boron-doped diamond may be viewed as a three-dimensional analog of $MgB_2$, which can be understood qualitatively in terms of a phonon mechanism.[34] Similarly, the temperature dependence of resistivity of the film under several values of magnetic fields up to 3 T for metallic example has also been measure in Fig. 3c, where the result of magnetic fields at -1.0 T and 1.0 T is shown in Supplementary Fig. S6. The field dependence of the onsets of $T_C$ is plotted in Fig. 3d, where the red dashed line is a linear fit to the onset values of $T_C$ over temperature which can be also expressed as $H$ = -1.5×$K$ + 4.8, where the $T_C$ onset can approach the value of 4.8 T at absolutely zero kelvin. The similarity between the metallic state and superconductive state as a function of temperature points to that the metallic state should also be caused the localization of bosons similar with the superconductive state. The co-doping of nitrogen and boron should bring the possibility of localization of bosons.[32]

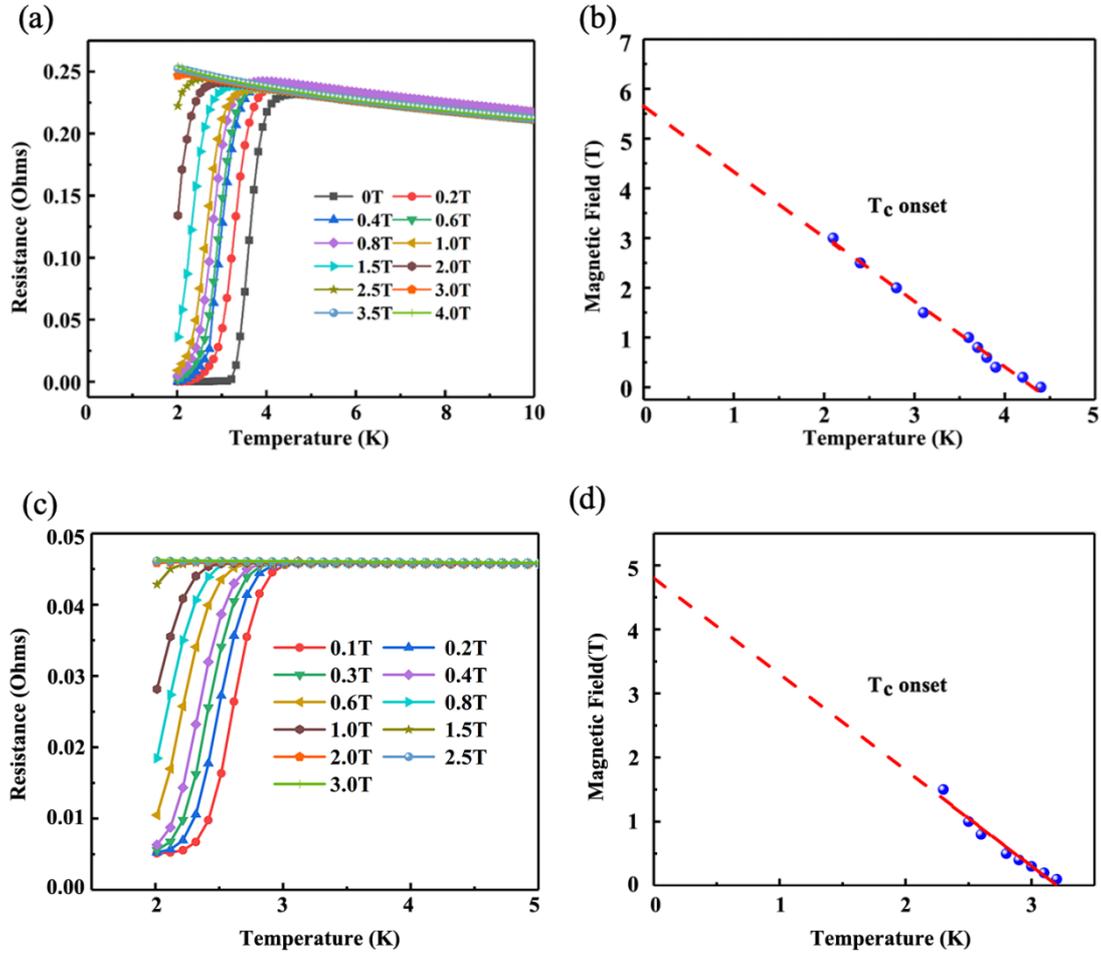

**Fig. 3 | Temperature dependence of resistivity under different magnetic fields for superconducting and metallic state samples. a,** the temperature dependence of resistivity under several values of magnetic fields from 0 T to 4 T for superconducting sample. **b,** field dependence of the onsets of $T_C$, the function of the fitted curve is $H = -1.3 \times K + 5.7$. **c,** the temperature dependence of resistivity under several values of magnetic fields from 0.1 T to 3 T for metallic sample. **d,** field dependence of the onsets of $T_C$, the function of the fitted curve is $H = -1.5 \times K + 4.8$.

Similar with diamond, graphene is carbon atom constructed material and graphene itself is not a superconductor due to the absence of electron-electron screening and subsequently weak electron-phonon interaction as the electron density of states (DOS) near the Dirac point is very low.[35] Graphene can be superconductive by placing it over superconductive substrate via the proximity effect.[36] On the other hand, twisted graphene can be superconductive due to the enhanced DOS as a result of flat band.[4-6]

Due to semiconducting nature of diamond, the graphene/diamond heterostructure should have different electrical characteristic as a result of high DOS in the heavily co-doped diamond. We have fabricated boron doped diamond named as sample D011515 with two times of nitrogen doping concentration compared with the samples listed in Table 1. The CVD grown graphene is transferred to diamond surface by wet method forming graphene/diamond heterostructure. As shown in Supplementary Fig. S7, TEM images indicate that the graphene/diamond have good interfacial contact quality. Besides, the EDS tests inside the diamond show that the diamond have uniform distribution of elements of boron and nitrogen in the co-doped diamond (Supplementary Fig. S8). In Fig. 4a, the resistivity of boron and nitrogen co-doped diamond heterostructure tends to decrease around 50 K, and it finally drop to zero below 2 K, as revealed by Fig. 4b. Fig. 4c shows the resistivity measurements of the graphene/diamond heterostructure and Fig. 4d shows the resistivity values below 10 K. From Fig. 4c and Fig. 4d, it is very interesting that the resistivity of graphene can drop to a negative value and finally reach zero resistance at 2K. This behavior is abnormal and repeatable, which could be ascribed to the specular Andreev reflection and exciton mediated superconductivity in graphene/superconductor junctions.[37] Actually, in 1970s, there is a report of electron pairing through the effect of exciton in metal-semiconductor Schottky diode.[38] Recently the novel mechanism of superconductivity has been proposed by electron-exciton and electron-polariton interactions.[39] The exciton binding energy of diamond is around 80 meV, which is quite large for initiating interlayer excitons with a large binding energy.[40,41] The interaction between charge-neutral excitons and free carriers is similar with the Fermi-polaron responses.[42] It is suggested that graphene might acquire strong superconducting properties below $T_C$ due to the exciton-mediated mechanism, as opposed to the acoustic phonon-mediated superconductivity.[31,43] Herein, in sample D011515, the nitrogen and boron atoms incorporated have a high doping concentration, thus it is possible to bring many excitons in the co-doped diamond. Putting graphene over diamond can facilitate the interaction between the holes in graphene and exciton in diamond or interlayer exciton formed in graphene/diamond heterostructure. In Fig.4c, the resistivity already begins to decay at 27 K, which may point to the freeze of exciton and the strong interaction between the holes in graphene and the exciton in heavily boron and nitrogen co-doped diamond. Compared with Fig. 4c, the drop of resistivity is very sharp in Fig.4d, which may mean that there is a strong interaction between the interlayer van der Waals excitons formed in graphene/diamond heterostructure and holes in graphene layer. Moreover, the Andreev reflection at the interface between the graphene/diamond brings many electron/hole pairs and the excitons can mediate the holes in graphene and diamond, which brings the superconductivity. Skopelitis *et al.* proposed that the interplay between phonon and exciton in hybrid semiconductor-superconductor may enhance the superconductivity behavior.[44] Herein, comparing Fig. 4d and Fig. 4b, the superconductivity transition range has become sharper while putting graphene onto the heavily boron and nitrogen co-doped diamond. In fact, the exciton-electron interaction has been observed in 2D materials under light illumination.[45] Under the framework of exciton-mediated superconductivity, the drag between Bose-condensed polaritons and electrons is allowed, which may cause the non-zero super current in the graphene/diamond heterostructure. Actually, Aminov *et al.* predicted the non-dissipative drag could be strong enough to be observable as induction of a supercurrent in the electronic layer by a flow of polariton Bose condensate.[46] The non-zero super current of graphene/diamond heterostructure may also be caused by the electric field induced symmetry breaking.[47]

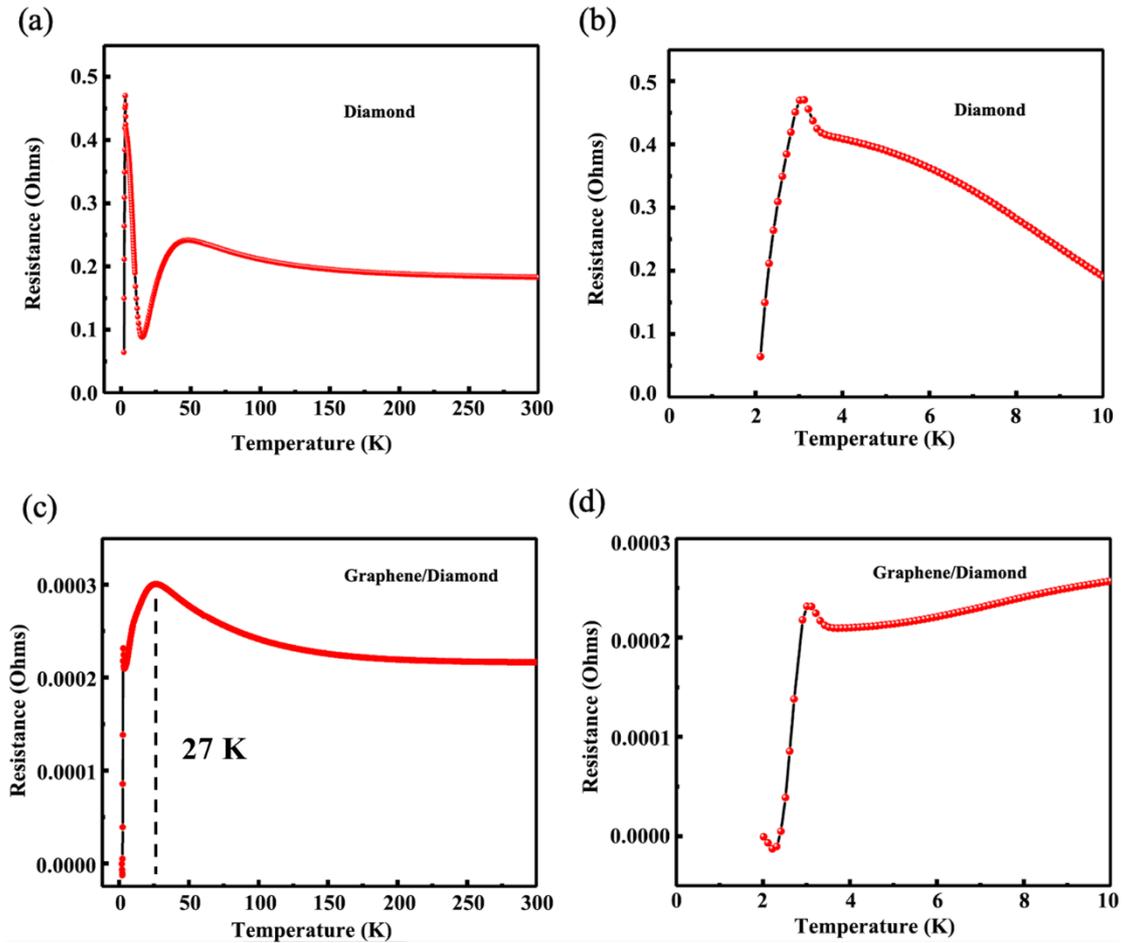

**Fig. 4 | Low temperature measurements of the conductivity of graphene/diamond heterostructure.** The dependence of diamond resistivity on temperature from 300 K to 2 K (**a**) and 10 K to 2 K (**b**). The dependence of the resistivity of graphene/diamond heterostructure on temperature from 300 K to 2 K (**c**) and 10 K to 2 K (**d**).

**Conclusion**

We have realized the metallic state in boron and nitrogen co-doped diamond. Through comparing the low temperature resistivity measurements and Hall effect results, it is revealed that the hole mobility plays an important role on deciding whether the sample is superconductive or metallic, where the larger mobility sample tends to be superconductive while the lower one tends to be metallic. Moreover, through increasing the nitrogen doping concentration, we find the sample can have a phase transition around 50 K, which may be caused by the exciton mediated copper pair formation. Through putting graphene onto heavily nitrogen and boron doped diamond, the resistivity can drop to a negative value before going to zero, which should point to the exciton mediated superconductivity in graphene contributed by the exciton inside the bulk diamond. In summary, this work will intrigue many studies over the metallic behavior in heavily co-doped semiconductor as well as the possible exciton mediated superconductivity both benefiting from the semiconductor nature of the diamond.

# Methods

## Materials fabrication and characterization

Single crystal diamond is grown by MPCVD using the $CH_4$ and Hydrogen as the main source, and a doped source of borane ($BH_3$) and Nitrogen ($N_2$) are added to form a mixed gas for growth. The samples of D121313, D010708, D120709 and D020405 were grown on (100) diamond seed under the 20 sccm flow of $BH_3$ and 5 sccm flow of $N_2$. The sample D011515 was grown on (100) diamond under the 20 sccm flow of $BH_3$ and 10 sccm flow of $N_2$. After growth, the single-crystal doped diamond was sequentially sonicated with acetone, isopropanol, ethanol and deionized water for 5 minutes to remove residues from the surface. And then, blow dry with a $N_2$ air gun. The laser Raman spectrometer (inVia-Reflex) is used for Raman testing of diamonds, using a laser light source of 532nm to scan the spectrum from 1000 $cm^{-1}$ to 2000 $cm^{-1}$. Diamond samples were measured using XRD (bruker D8 ADVACNCE) and scanned at 0.02° in the range of 0° to 130°. The Hall effect test instrument (Nanometrics HL5500) was used to determine the carrier concentration, resistivity and Hall mobility of single crystal diamond. The samples used for diamond TEM characterization were prepared by a dual-beam focused ion beam microfabrication instrument (FEI, Quata 3D FEG) to realize the nanoscale processing of diamond samples for TEM characterization. TEM images were obtained by spherical aberration corrected transmission electron microscopy (FEI Titan G2 60-300) to observe the atomic structure arrangement and element distribution in the body of single crystal diamond.

## Superconductive measurements

The surface of diamond ohmic contact is grown by magnetron sputtering with an Au electrode of 100 nm (the spacing between the electrodes was 1.0 mm) for the measurement of low-temperature electrical properties. As for graphene/diamond heterostructure, the single-crystal graphene on the copper foil grown by CVD was used to form a heterojunction structure with doped diamond by means of wet transfer. Similarly, magnetron sputtering was used for low-temperature electrical testing of gold electrodes with a surface growth gap of 1.0 mm and a thickness of 100 nm. A low-temperature curing silver paste was applied to the surface of the gold electrode to connect the gold wire (diameter was about 50 μm) to the measuring base of the low-temperature electrical test equipment. Superconductivity properties were measured using a low-temperature magnetic field test and specimen preparation system (QUANTUMDESIGN, PPMS-9) with a test range from 2 K to 300 K and the test intervals of 0.1 K. Diamagnetic testing and variable magnetic field resistance testing are performed using a low-temperature magnetic field test and specimen preparation system (QUANTUMDESIGN, MPMS-XL-5) with a magnetic field variation range of 0 to 4 T.

**Acknowledgements**

S.S.L. thanks the support from the National Natural Science Foundation of China (No. 51202216, 51551203, and 61774135), Youth Talent from the Central Organization Department.

**Conflict of Interest**

The authors declare that they have no known competing financial interests or personal relationships that could have appeared to influence the work reported in this paper.

**Author Contributions**

S.L. designed and carried out the experiments and processed the data, discussed the results, wrote the paper and conceived the study. X.Y. carried out the characterizations and processed the data. M.Y., H.Z. and J.G. helped the experiment and discussed the results.